\documentstyle[epsf,aasms4]{article}

\def\arcsec{\hbox{$^{\prime\prime}$}}
\def\sqas{\hbox{$\Box^{\prime\prime}$}}
\def\sqdeg{\hbox{$\Box^\circ$}}
 
\def\Msun{\hbox{$M_\odot$}}

\def\Hnull{\hbox{$H_0$}}

\def\cm2{\hbox{$cm^2$}}
\def\cm3{\hbox{$cm^3$}}
\def\cmm2{\hbox{$cm^{-2}$}}
\def\cmm3{\hbox{$cm^{-3}$}}

\begin{document}

\title{Faint galaxies around quasars at $z=1$ and gravitational 
lensing of distant objects}

\author {J.W. Fried}

\affil{
Max-Planck-Institut f\"ur Astronomie, K\"onigstuhl 17\\
D-69117 Heidelberg, Federal Republic of Germany}
 
\authoremail{fried@mpia-hd.mpg.de}

\begin{abstract}

Very deep imaging data of three optically luminous radio-loud quasars with 
redshifts between $z=0.9$ and $z=1.36$  are presented. The data are 
complete for galaxies down to $R=26$. There is no evidence for 
excess numbers of galaxies around the quasars; foreground galaxy
clusters are excluded by the data as well as clusters with richness 
classes greater than 1 associated with the quasars.
We find clear evidence for gravitational lensing for two quasars. 
It is further shown that due to the high surface density of galaxies 
all distant ($z>1$) objects are moderately affected by gravitational
lensing; the amplification factors are estimated to be in the range $1.1-1.5$.

\end {abstract}

\section {Introduction}

Galaxies around low redshift ($z = 0.3-0.6$) quasars have been studied 
extensively by various authors (Gehren et al. 1984, Green and Yee 1988, 
 Hintzen et al. 1991, Ellingson, Yee and Green 1991, 
Fried 1991); there is general agreement that 
low redshift  radio-loud quasars are found in moderately rich 
galaxy clusters with richness classes $0-1$. Radio-quiet QSOs are 
usually found in less dense environments. Green and Yee (1988) and
Fried (1991) have found evidence for an evolution of the environment 
between redshift $z=0.3$ and $z=0.6$ in the sense that the qso-galaxy
covariance amplitude is larger at $z=0.6$.

At redshifts $z \approx 1$ associations between different types of 
AGN and galaxies have been sought by Tyson 1986, 
Hintzen, Romanishin and Valdes 1991, Fried, Stickel and K\"uhr 1993, 
Hutchings, Crampton and Persram 1993, 
Hutchings, Crampton and Johnson 1995, Boyle and Couch 1993, and 
Benitez et al. 1995. 
These works are based on imaging data with limiting magnitudes around 
$23.^{mag}$. Since a galaxy with $M^*=-21$ appears as faint as 
$m=24.35$ at redshift $z=1$ (choosing $\Hnull = 75\, km\, s^{-1}\,Mpc^{-1}$ , 
$q_0 = 0.5$ and applying an R-band K-correction of 2 mag according 
to Metcalfe et al. (1991)) this implies that galaxies associated 
with QSOs at $z \ge 1$ can be detected in such data  only if they 
are overluminous by $1-3\, mag$. 
Furthermore, only the brightest members of possible associated clusters 
can be detected. The results of these studies are discrepant, there are 
claims both for and against excess numbers of galaxies 
(a detailed discussion is given below). Since redshifts of these galaxies 
are not available, the excess numbers of galaxies are interpreted either 
as true physical  association of these galaxies with the QSOs or as 
foreground galaxies acting as gravitational lenses. 

In this paper we present data on the fields around 3 quasars. The
quasars were selected from the catalogue of Veron \& Veron (1993); they 
are all radio-loud and have redshifts $z=0.901$ (PKS 2216-03), 
$z=1.066$ (PKS 2356+196) and $z=1.36$ (OT 566). The data are complete 
for galaxies to $R=26$ and thus allow detection of galaxies at the 
redshifts of the  quasars down to $M^*+2$. 
The data were originally taken to search for a possible galaxy excess
around the quasars; this is discussed in section 3.1. It was also 
found that the data have implications for gravitational lensing
(section 3.2). In section 3.3 it is shown that the size of the absorber 
in the line of sight to PKS 2216-03 may have been severely overestimated.

\section{Observations and data reduction}

The direct images presented here were obtained at the prime focus of
the 3.5 m telescope on Calar Alto, Spain. All observations were made 
in the R-filter band. A CCD with 640x1024 pixels of $15 \mu m$  was used as 
detector. The long exposure times which are necessary to reach 
faint light levels were broken into many shorter exposures of 500 sec each, 
with the telescope moved  between the exposures by several steps of 
$4 \arcsec\ $ each in right ascension and/or declination; 
flatfields for each night were 
constructed from an unregistered  median of the individual frames.
The frames were debiased, flatfielded and then registered. To correct 
cosmic ray events, the median and sigma for each pixel were 
calculated for 6-8 registered frames; those pixels which deviated more 
than $\approx 3\, \sigma$ from the smoothed median were replaced 
by the median.  This procedure removes cosmic ray events very
effectively even in the flanks of stellar images. These cosmic ray 
free frames were then summed  to construct the final coadded images. 
It should be emphasized that the median is used only 
for constructing a flatfield and for correction of cosmic events,
since Steidel and Hamilton (1993) have shown that the median operation 
changes the true intensity distribution of objects close to the sky level. 
The total integration times are between 13.5 ksec and 26.0 ksec; 
the $1\sigma$ surface brightness limits are between 
$\mu_R = 29.6 mag\,/ \sqas$ and $\mu_R = 30.1 mag\,/\sqas$. 
The central parts of the three fields are shown in figure 1.

Since the weather was not photometric during the whole observing run, 
frames of the three objects obtained during photometric conditions 
were calibrated with observations of standard stars, and 
these calibrations were used to scale the final coadded images. 
Photometric errors due to atmospheric instabilities should be less 
than 0.1 mag. The seeing was acceptable throughout the whole 
observing run; the coadded images have PSFs with FWHM $1.1 - 1.2\arcsec\ $.

\begin{figure*}[h] \label{images}

\epsffile{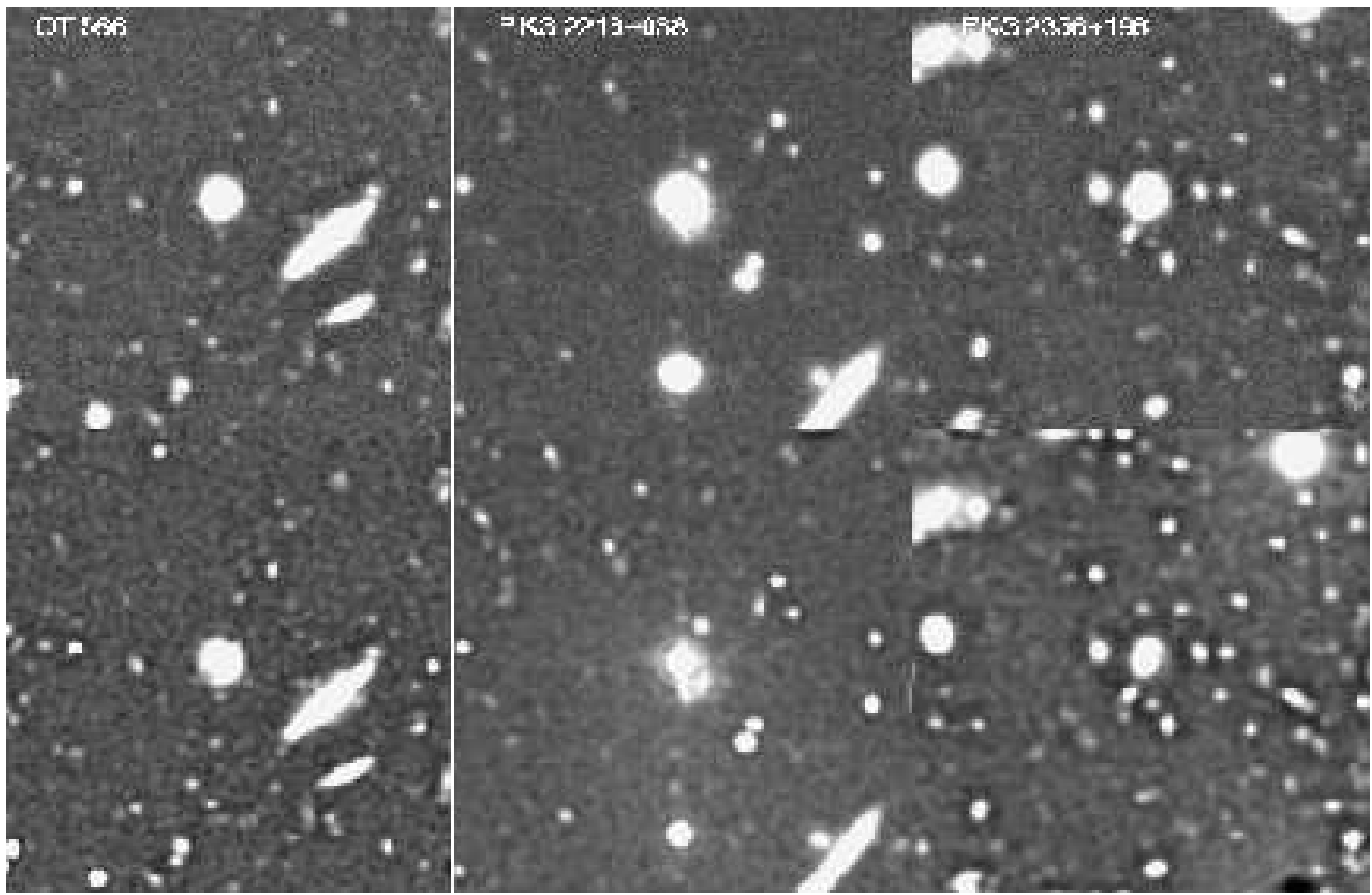}

\caption{Fields around three quasars. The quasars are in the field
centers; the fields are 64 $\arcsec $ square. The upper panels show 
the original, the lower panels the deconvolved data. North is to the top, 
east to the left.} 
\end{figure*}

The detection of objects and separation into stars, galaxies and 
image defects was done using the FOCAS software (Valdes 1982). 
The completeness limit can easily be determined from differential number 
counts (fig.2). These number counts are {\em not} corrected for 
incompleteness due to diminishing detection probability of the 
objects near the faint end and thus demonstrate clearly that the data
are complete for galaxies down to $26^{th}$ magnitude. 
The data are well fitted by a linear  relation 
$log N(m) = slope \cdot R+constant$. 
The slope and constant are 
$0.31 \pm 0.01$ and $ -2.91 \pm 0.31$ for OT566, 
$0.36 \pm 0.08$ and $ -4.28 \pm 0.18$ for PKS2216-038,
and  $0.29 \pm 0.01$ and $-2.40 \pm 0.23$ for PKS2356-196 in very good
agreement with other determinations reaching $R \ge 26$ 
(Tyson 1988, Steidel and Hamilton 1993, Metcalfe et al.1995, 
Smail et al. 1995).

\begin{figure}[htb] \label{numcounts}
\centerline{\hbox{\epsffile{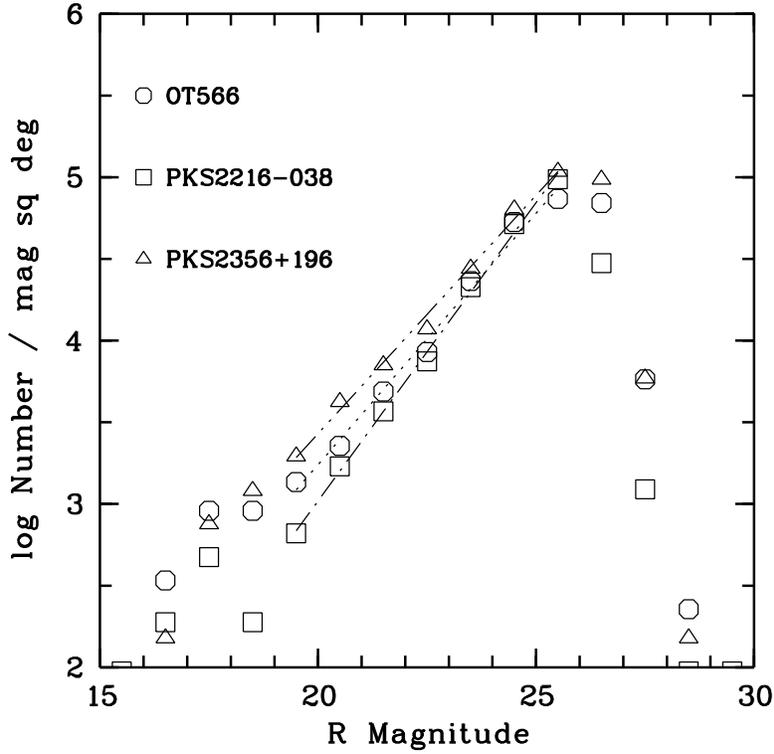}}}
\caption[ ]{Differential galaxy number counts in the fields of the quasars.
The straight lines show the fits to the data in the range $R=18-26$}
\end{figure}

\section{Results}

\subsection{Excess of galaxies around the quasars?}

A possible excess of galaxies around the quasars is most easily
detected on plots of galaxy surface density (galaxies per square
degree) as a function of projected distance from the quasars. 
Foreground clusters of galaxies should show up in the $R = 18-23$ 
subsample, whereas clusters of galaxies associated with the quasars 
should show up in the $R = 23-26$ subsample because the apparent 
magnitudes would be in this range (see introduction). From figure 3 
it is evident that the galaxy surface density is constant, i.e. there
is no evidence for a galaxy excess in either field. In table 1, the
number  of galaxies counted within  various radii
from the quasars is compared to the number expected from
background. Since there is no radial gradient in the data, the
background  has been taken as the mean over each individual field.
However, while fig.3 and table 1 suggest that there is no galaxy excess 
in any of the 3 fieds, statistical fluctuations  in the background could 
exceed the number of galaxies expected for poor clusters. Table 2 lists
the ratio of expected numbers of galaxies within certain radii for Abell clusters  
of richness classes  0 - 3 and the $1\sigma$ Poisson fluctuations in the
background counts. The number of galaxies expected within a given
radius was integrated for a King profile with core radius $r_c = 250
h_{50}^{-1} kpc $ (Bahcall 1975) and the total numbers of cluster members
were taken as the mean of the population numbers for each richness class 
given by Abell 1958. All counts  were taken as the mean of the 3 fields.
Table 2 shows that foreground clusters can be excluded on a $3\sigma$ 
level, clusters associated with the quasars only for richness classes
greater than 1.

\begin{figure*}[htb] \label{ring}

\centerline{\hbox{\epsffile{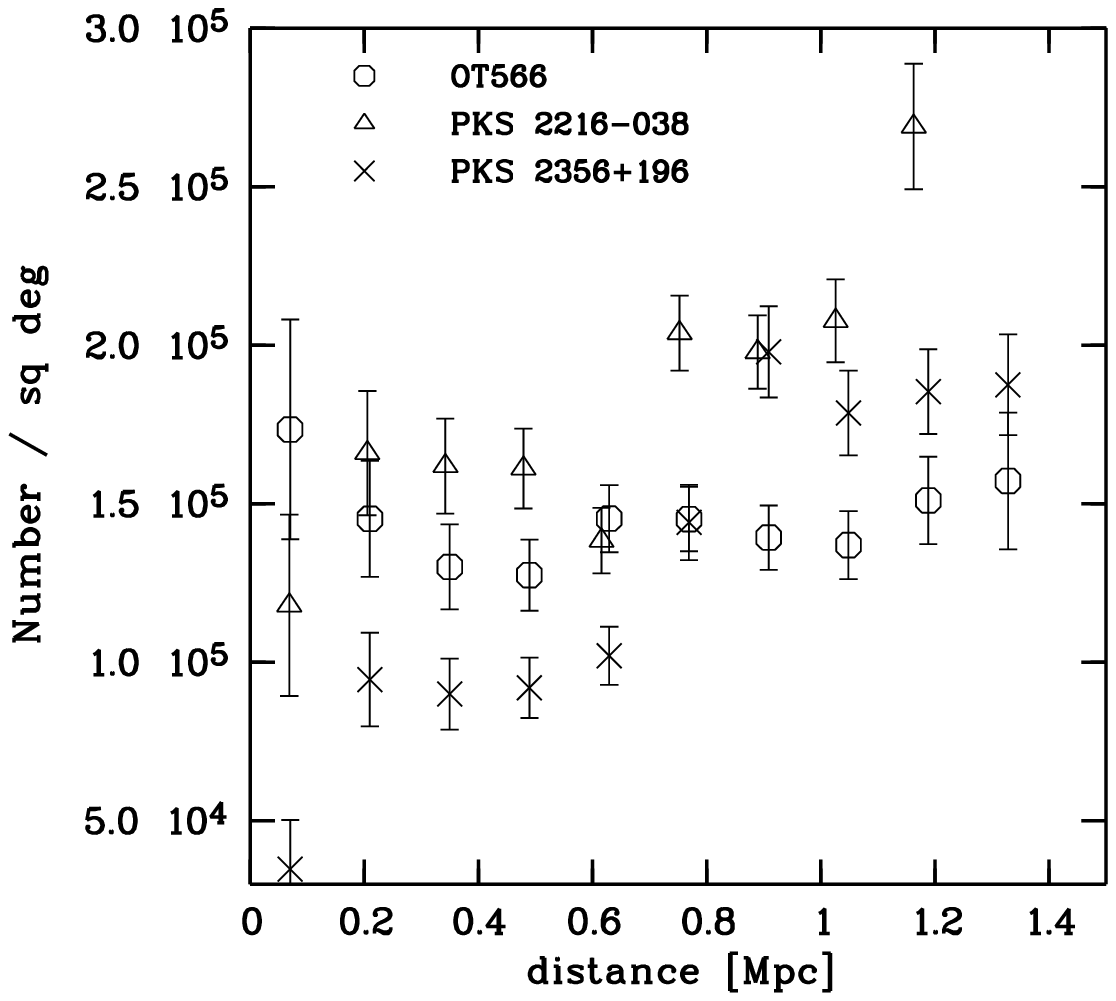}}}
\centerline{\hbox{\epsffile{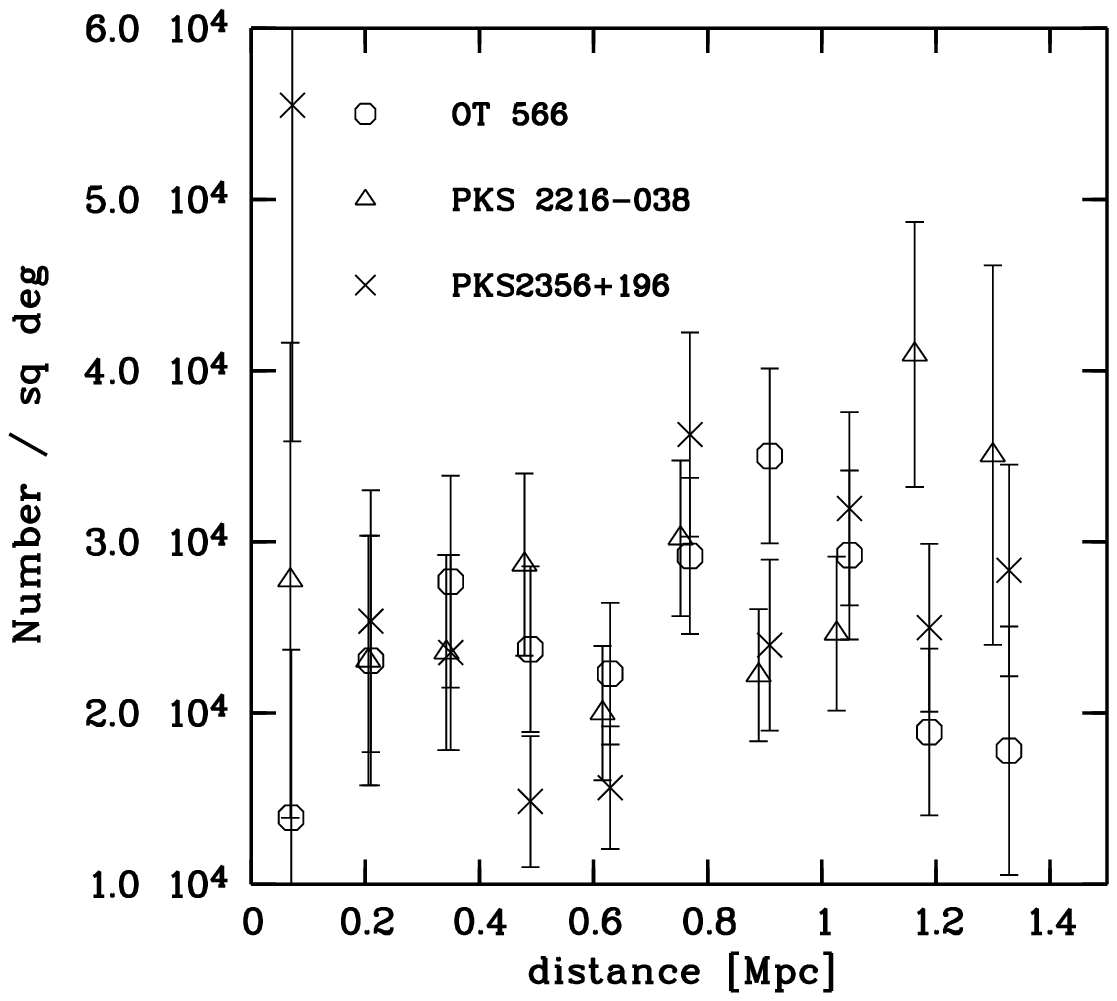}}}
\caption[ ]{Galaxy surface densities (galaxies per square degree) as a
function of projected radial distance from the quasars. At $z=1$,
100 kpc correspond to $\approx 18 \arcsec$. The lower panel 
contains galaxies with $R=18-23$, the upper panel galaxies with 
$R=23-26$. The error bars are computed from Poisson statistics ($1\sigma$)} 
\end{figure*}

\begin{table*} \label{cluster1}
\caption [ ]{Counted/expected numbers of galaxies from background with
$R=18-23$  and $R=23-26$ within the given radii from the quasars.}

\begin{flushleft}
\begin{tabular}{|c|ccc|ccc|} \hline 
         &         & ${\rm R}=18-23$ &      &     & ${\rm R}=23-26$ &\\ \hline
quasar       & 100 kpc & 250 kpc & 500 kpc & 100 kpc & 250 kpc & 500 kpc \\ \hline
OT 566       & 1/1.3   &   8/8.3 & 33/33.1 & 14/11.1 & 73/69.3& 256/277.1 \\
PKS 2216-038 & 3/1.1   &   8/6.6 & 29/26.4 & 10/13.1 & 77/81.7& 337/327.0 \\  
PKS 2356+196 & 3/1.9   &  18/12.2 & 49/48.7& 5/14.8  & 81/92.6& 333/370.4 \\ \hline    
\end{tabular}
\end{flushleft}
\end{table*}

\begin{table*}[here] \label{cluster2}
\caption[]{Ratio of expected number of galaxies within given radii for
galaxy clusters of given richness class to $1 \sigma$ Poisson
fluctuations in the background for the two subsamples with  
$R=18-23$ and $R=23-26$}
\begin{tabular}{|c|c|cccc|c|cccc|} \hline
\multicolumn{6}{|c}{$R=18-23$}&\multicolumn{5}{|c|}{$R=23-26$}\\ \hline
\multicolumn{2}{|c}{ }& \multicolumn{4}{c|}{richness class} & 
\multicolumn{1}{c}{} &   \multicolumn{4}{c|}{richness class}  \\ \hline
r[kpc] &      back    &  0  & 1    & 2   & 3 &   back        &  0  & 1    & 2   & 3 \\ \hline
100    & $1.43 \pm 1.2$ & 0.7 & 1    & 1.7 & 2.6& $13\pm3.6$ &2.1  & 2.9 & 5 & 7.9  \\
250    & $9\pm 3 $ & 1   & 1.7  & 2.7 & 5.5& $81 \pm 9$  & 3 & 5& 16.3 & 16.7\\
500    & $36 \pm 6$ & 1.1 & 1.8  & 2.9 & 4.6&  $325 \pm 18$ & 3.3 & 5.5 & 8.8 & 13.8\\ \hline
\end{tabular}
\end{table*}

\subsubsection{Comparison with other work}

The quasars studied here are all radio-loud and optically luminous, 
the mean absolute visual magnitude for our small sample being 
$<M_V> = -25.9$. Samples with similar properties have been studied by 
Tyson (1986) and Hintzen, Romanishin and Valdes (1991).

Tyson (1986) found a significant excess of bright galaxies with $m< 21$  
within $30 \arcsec$ of the quasars, which he believed to be due 
to enhanced luminosity evolution of galaxies near the quasars. However 
his data are not easily comparable to other work since they 
were measured in white light and the galaxy excess was 
detected only after considerable correction for decreased detection 
probability of galaxies close to a bright point source. 
Hintzen, Romanishin and Valdes (1991) could not confirm Tyson's 
results, but found an excess of fainter galaxies ( $R<23$) closer to 
the quasars ($ < 15 \arcsec$ ). They found 33.8 galaxies within $15
\arcsec$ from the 16 quasars in their sample, compared to 19.3
galaxies expected from background counts which were determined 
in the outer regions of the fields. This excess is significant on 
the $3 \sigma$ level using Poisson statistics. However, the fluctuations 
in galaxy counts are larger than given by Poisson statistics by about 
a factor of 2 in fields of few arcmin size (Metcalfe et al. 1991,
Tyson 1990). Thus the significance of the galaxy excess may be 
overestimated. Hintzen, Romanishin and Valdes (1991), too, interpreted 
the excess as being due to luminosity evolution by few magnitudes of the 
galaxies near the quasars. 

If the unified scheme for AGNs is correct, then BL Lac objects 
owe their special properties to aspect angle; 
therefore they should be found in environments similar to those of
radio galaxies and quasars. Fried, Stickel and K\"uhr (1993) analysed 
the environments of 1 Jy radio selected BL Lac objects. For 
redshifts $z \le 0.6$ the environments are similar to those of 
radio-loud quasars, i.e. clusters of richness class $0-1$ with denser 
environment at $z=0.6$. The subsample with $<z>=1$ consisted of 
5 luminous ($<M_V>=-25.4$) objects; no clustering was found, but there was 
evidence for gravitational lensing for 2 objects. 

The sample of Hutchings, Crampton and Persram (1993) 
and Hutchings, Crampton and Johnson (1995) contains 14 QSOs which 
are optically fainter with $<M_V> = -24.4$. The authors report an 
excess of galaxies over the background  in 100 arcsec (568 kpc) 
subfields near the QSOs on the $2.5 - 3 \sigma$ level; they 
interpret their data as evidence that QSOs are located in compact 
groups of starbursting galaxies.  Three of the objects are radio-loud; 
the authors found no significant difference between radio-quiet and 
radio-loud objects. 

A sample of 27 slightly fainter $(<M_R> = -23.4)$ radio-quiet QSOs 
has been analysed by Boyle and Couch (1993); there was no significant 
excess of galaxies with $R\le 23.0$ around the QSOs.

At fainter optical magnitudes, Benitez et al.(1995) found an excess of 
bright ($19.5 < R < 22$) galaxies around 5 radio galaxies at 
the 99.4\% confidence level. Since this excess was not present 
for fainter ($22 < R < 23.5$) galaxies, the authors concluded that 
the galaxy excess is due to foreground objects and thus related to 
gravitational lensing.

This compilation shows that there is currently no consensus about an 
excess of galaxies around quasars. One might suspect that part of the
discrepancies between the different studies  might be due to luminosity 
effects both in the optical and radio regimes. However, as our 
compilation shows, the results are contradictory  even when 
similar samples are compared. Our data are deeper by about 3 mag than 
the cited studies, but   in the 3 cases
studied here we find no evidence for a galaxy excess. 
There is also no consensus about the interpretation of a galaxy excess; 
since the redshifts of the galaxies around the quasars are not known, 
any detected excess can be interpreted either as true physical association 
or lensing of foreground objects.

\subsection{Two lens candidates}

Two of the three quasar images (PKS 2216-038 and PKS 2356+196)  are definitely 
non-stellar (see fig.1). The MIDAS implemenation of the Lucy-Richardson 
deconvolution algorithm was used to reveal the underlying structure. 
Images of stars, which had  neighboring objects removed by interpolation, 
were  used as PSFs. Already after very few iterations the deconvolution 
shows that the images of faint galaxies are superposed onto the 
images of the quasars PKS 2216-038 and PKS 2356+196; the final result of the
deconvolution is virtually independent of the number of iterations. 
Table 3 lists the relevant data for the  galaxies next to the 
quasars.

\begin{table} \label{galaxies}
\caption[]{ Photometry and positions of galaxies closest to the line of
sight to PKS2216-038 and PKS2356+196}

\begin{tabular}{lccc} 
		   &  distance &  pos. angle & R  \\
	           & $\arcsec$ &  degree     & mag\\  \hline
PKS 2216-038        &  5.6      &  330           &21.9\\
	 	   &  4.2      &   185          &21.5\\
                   &  4.1      &   220          &21.7\\
PKS 2356+196        &  2.1      &  340           & 20.6\\
                   &  5.0      &   162 		&21.6\\
                   &  6.0      &   192            & 23.7\\
                   &  7.6       &  278            & 21.9\\ \hline
\end{tabular}
\end{table}

These galaxies appear very similar to the ones seen 
in the fields of the quasars. From redshift surveys it is known 
that galaxies with $B \approx 24$ have median redshift $z=0.46$ 
(Glazebrook et al. 1995).  Therefore  it appears reasonable to assume 
that the galaxies close to lines of sight to the quasars, too, are at $z
\approx 0.5$; if so they are foreground to the quasars and amplify 
the light from the quasars by gravitational lensing. 

Since redshifts and velocity dispersions of these galaxies are 
unknown, only crude estimates of the amount of lensing can be made: 
assuming $z_{lens}=0.5$ and velocity dispersions for the lenses 
$\sigma_l = 200\, km\,s^{-1} $ and $\sigma_l = 300\, km\, s^{-1} $, 
the formula given by Turner, Ostriker and Gott (1984) gives 
amplification factors 1.2 and 2, respectively. The required  masses
are a few $10^{11} \Msun$ and for a typical magnitude $R=21.5$ (see
table 2) one obtains an absolute magnitude $M_R=-21$ using the chosen
cosmology and a K-dimming and evolution correction of $1 mag$ 
according to Metcalfe et al. (1991). Therefore, these possible lenses
would have masses and absolute magnitudes like typical bright galaxies.

It should be noted that PKS 2216-038 is included in the complete 
1 Jy sample of radio sources; since its radio flux at 6cm wavelength 
is  1.5 Jy, this object may have entered the sample because it is 
lensed. Further evidence for lensing in this object has been given 
by Bartelmann, Schneider and Hasinger (1994); they have found a 
correlation between X-ray emission from foreground galaxy clusters 
and high-redshift radio-loud quasars from the 1 Jy-sample. 
PKS2216-038 is among the objects which show the highest correlation.

\subsection{Lensing of distant objects}

The integral surface density  of galaxies $R \le 26.$ 
is $1.93\,10^5\,/ \sqdeg$ (mean of the three fields); this results in a 
projected mean distance between galaxies of only $8.2 \arcsec$. Thus it 
is evident that many lines of sight to background (i.e. $z \ge 1$) 
objects pass close enough by a galaxy that they are affected by lensing.

Fig.4 shows the probability distribution of the
amplification  factor for the field of OT566 (the other fields give
practically identical results). This was computed by putting random
points into the field, and then calculating the distance to the nearest
neighbor and the resulting amplification with the formula given by 
Turner, Ostriker and Gott (1984) for velocity dispersions of 
$200\, km \, s^{-1}$ and  $300\, km \, s^{-1} $, which correspond to mean
galaxy masses of $8.1 \, 10^{11} \Msun$ and $1.8 \, 10^{12} \Msun$, 
respectively. These masses are in good agreement with recent dynamical
mass estimates for nearby galaxies (Zaritsky and White 1994) and for 
faint galaxies within the magnitude range $20 \le r \le 23$ 
(Brainerd, Blandford and Smail, preprint). The redshifts of the lenses
were distributed randomly between 0 and 1, and for the source we have
taken $z_{source}=1.2$.  The assumed redshifts affect the result only 
weakly over a very wide range of redshifts. The diagram shows that the 
bulk of amplification factors relative to an 'empty' line of sight is between 
1.1 and 1.5, the effect of lensing thus is moderate. From the width of 
the distribution of the amplification factors it is also obvious that lensing 
imposes a scatter on the observed luminosity function of distant
objects. It must be emphasized that the assumptions on the velocity
dispersions of the galaxies made for these
estimations affect the amplification factor only quantitavely ; the
qualitative conclusion, that distant objects must be lensed by
foreground galaxies, is based  purely on the observed galaxy surface 
density. We are therefore {\em forced} to conclude that distant 
($z \ge 1$) objects are lensed by foreground galaxies. 
On theoretical grounds this was anticipated by Press and Gunn (1973), 
who had concluded that all distant objects in a $\Omega=1$ cosmology 
are lensed.

\begin{figure}[h] \label{ampli}
\centerline{\hbox{\epsffile{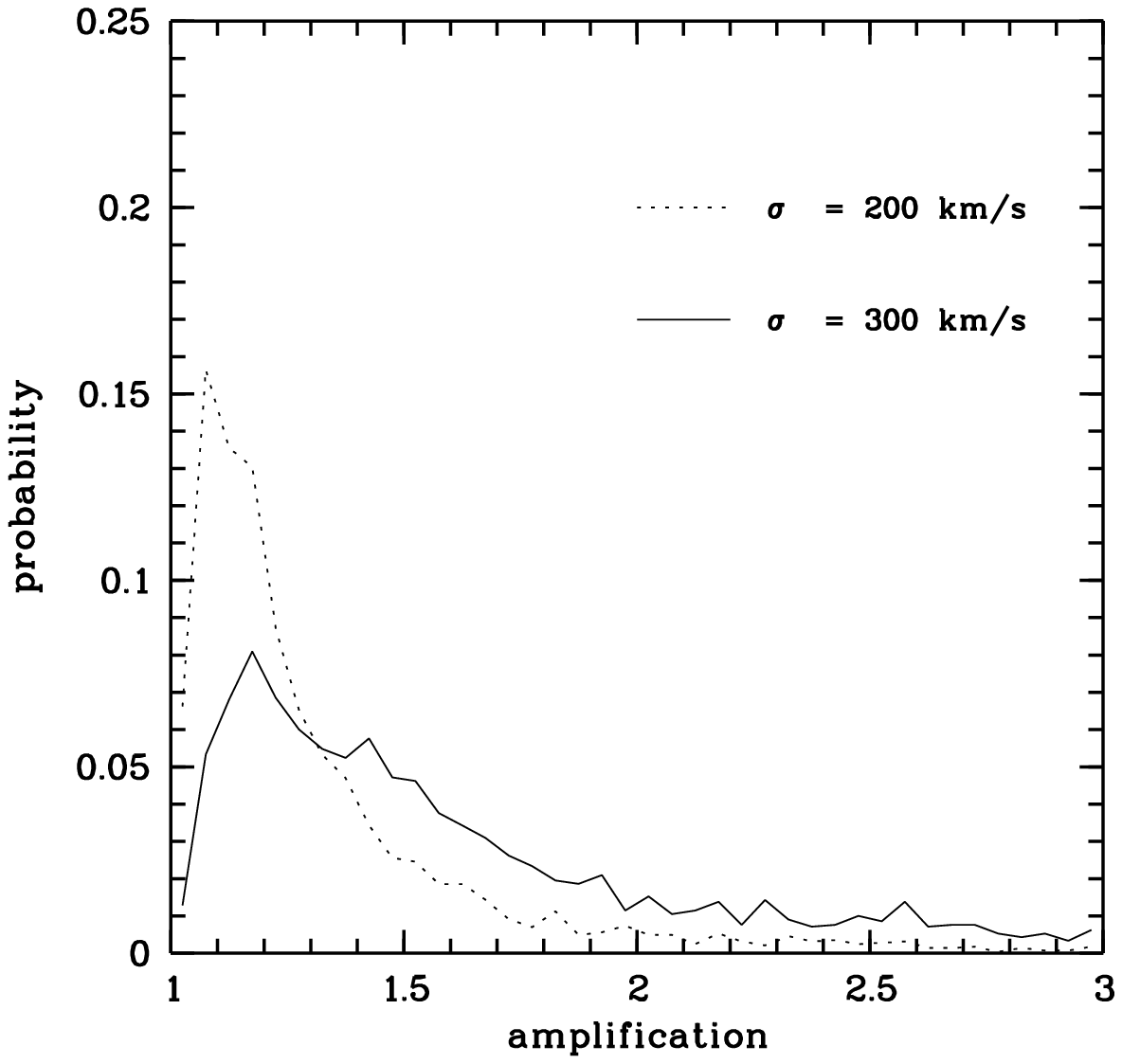}}}
\caption{Probablility distribution of the amplification factors  
of random background ($z=1$) objects by the galaxies in the field of OT566} 
\end{figure}

It is interesting to estimate the effects of microlensing. The optical depth 
$\tau_{\mu l}$ to microlensing is equal to  the ratio of the
surface density of the microlensing matter to the critical mass
density $\Sigma_{crit} = \frac{c^2}{4\pi G D}$
where $D = \frac{D_l D_{ls}}{D_s}$ and all symbols have their usual 
meaning. $\Sigma_{crit}$ can be calculated from the chosen cosmology 
and the assumed redshifts $z_{source} = 1$ and $z_{lens} = 0.5$. 
The surface mass density of the lensing matter can be  estimated for 
elliptical galaxies from $\Sigma_e = (M/L)\, I_e$ where the subscript e 
denotes the effective radius. Since $I_{tot} =  22.4 I_e r_e^2$ 
we obtain $\Sigma_e = \frac{M_{tot}}{ 22.4 r^2_e}$.
Taking $M=10^{12}\, \Msun$ and $r_e = 5\, kpc$ gives 
$\tau_e = 1/2.5 (M/10^{12} \Msun) (r_e/5kpc)^{-2} $.
For the random points which were put into the galaxy field of OT566
(see above), the mean distance to the nearest neighbor is about 
$4 \arcsec$ or $18.6\, kpc$; scaling $\tau_{\mu l}$ according to the $r^{1/4}$-law 
gives a mean optical depth to gravitational microlensing of $\tau=0.016$. 
This is well below unity, showing that microlensing by galaxies
with $R \le 26$ is not very important. An optical depth  $\tau_{\mu l}  
\approx 1$ would require a galaxy surface density about 15 times the
one measured to $R \le 26$. Now, the differential galaxy 
number counts rise continuously to $R=26$ with an increase in galaxy
surface density by about a factor 2.3 per mag. The counts of 
Smail et al. (1995) show that this continues 
to $R=27.5$ with no indication of a declining slope. If the counts 
would rise continuously to $R=29-30$,  
then the nearest neighbor distance $ \approx 5\, kpc$ and 
thereby $\tau_{\mu l} \approx 1$. Evidence that all distant objects 
are indeed microlensed has been given by Hawkins (1993);  he 
measured the  lightcurves of many QSOs and found that the origin
of variability of high-redshift QSOs ($z \ge 1$) is not instrinsic 
to the QSOs, but rather caused by microlensing. Since 
Hawkins and Veron (1993) had previously  shown that 
virtually  {\em all} QSOs are variable to some extent, these results 
present strong evidence that all distant objects are actually microlensed.

\subsection{Galaxy halos}

Accidentally the quasar PKS 2216-038 was also included in the sample of Le
Brun et al. (1993) who were searching for galaxies close to quasars with
MgII $\lambda\lambda 2798$ absorption lines in their spectra. 
Le Brun et al. (1993) report a possible detection of this absorption line at
a redshift $z_{abs}=0.202$ and identified a galaxy 34 arcsec away
from the quasar as absorber; this galaxy would have a large 
halo of $102 h_{75}^{-1}$ kpc radius. However, as fig. 1  shows
there are several galaxies within few arcsecs to the line of sight to 
the quasar. These galaxies have apparent magnitudes around $R=21.5$; at the 
absorption line redshift their absolute magnitudes would be $M = -18.6$ 
(neglecting K-correction and evolution, but both effects are small at 
such low redshifts). York et al.(1986) have shown that even 
low-luminosity galaxies such as the Magellanic Clouds can cause 
absorption lines in the spectra of distant objects. Therefore, 
the galaxies few arcsec from the quasar can well be the true
absorbers; if so, the size of the halo of the absorbing galaxy 
given by LeBrun et al. (1993) may be severely overestimated. 
Due to the high surface density of faint galaxies, cases like 
PKS 2216-038 may not be exceptional and may at least in some cases 
lead to severe overestimation of the sizes of galaxy halos.

\section{Summary}

We have presented very deep direct images of the fields of three quasars with
redshifts between $z=0.9$ and $z=1.36$. 

Counts of galaxies which are complete down to $R=26$ give no 
evidence for a galaxy excess around the quasars in any of the three fields. 
We find no evidence for foreground clusters; associated galaxy
clusters with richness classes larger than 1 are excluded. 

Two of the quasars have faint galaxies within few arcsec to their 
line of sight; since these galaxies are almost certainly foreground, 
the quasars are lensed with amplification factors probably between 
1.2 and 2. 

The high surface density of galaxies leads to the conclusion that 
virtually all distant $z \ge 1$ objects are affected by gravitational 
lensing. The bulk of amplification factors relative to an 'empty' beam are
estimated to be in the range 1.1 to 1.5.

The quasar PKS 2216-038 may also have an MgII absorption line at $z=0.2$ 
in its spectrum; if one of the galaxies detected in our image is the true
absorber, then only a moderately sized halo of 10 kpc radius is required in
contrast to claims of a huge 102 kpc halo. 

\acknowledgements{ Suggestions from G.Williger have improved the
manuscript. Discussions with J.Wambsganss have been very
helpful and are gratefully acknowledged.}

\end{document}